\documentclass[a4paper,10pt]{article}
\usepackage[utf8]{inputenc}
\usepackage{url}
\usepackage{amsmath}
\usepackage{amssymb}
\usepackage{pdflscape}
\usepackage{rotating}
\usepackage{float}
\usepackage{xcolor}

\usepackage{color}

\floatstyle{ruled}
\newfloat{algorithm}{tbp}{loa}
\providecommand{\algorithmname}{Algorithm}
\floatname{algorithm}{\protect\algorithmname}

\begin{document}

\begin{center}
{\Large \textbf{Bayesian Inference for Duplication-Mutation with Complementarity Network Models}}

\vspace{0.5cm}

BY  AJAY JASRA$^{1}$, ADAM PERSING$^{2}$, ALEXANDROS BESKOS$^{2}$, KARI HEINE$^{2}$, \& MARIA DE IORIO$^{2}$

{\footnotesize $^{1}$Department of Statistics \& Applied Probability, National University of Singapore, 6 Science Drive 2, Singapore, 117546, SG.}\\
{\footnotesize E-Mail:\,}\texttt{\emph{\footnotesize staja@nus.edu.sg}}\\
{\footnotesize $^{2}$Department of Statistical Science, University College London, 1-19 Torrington Place, London, WC1E 7HB, UK.}\\
{\footnotesize E-Mail:\,}\texttt{\emph{\footnotesize a.persing@ucl.ac.uk, a.beskos@ucl.ac.uk, k.heine@ucl.ac.uk, m.deiorio@ucl.ac.uk}}\\

\vspace{0.35cm}

\end{center}

\begin{abstract}
 We observe an undirected graph $G$ without multiple edges and self-loops, which is to represent a protein-protein interaction (PPI) network. We assume that $G$ evolved under the duplication-mutation with complementarity (DMC) model from a seed graph, $G_0$, and we also observe the binary forest $\Gamma$ that represents the duplication history of $G$. A posterior density for the DMC model parameters is established, and we outline a sampling strategy by which one can perform Bayesian inference; that sampling strategy employs a particle marginal Metropolis-Hastings (PMMH) algorithm. We test our methodology on numerical examples to demonstrate a high accuracy and precision in the inference of the DMC model's mutation and homodimerization parameters. 
\end{abstract}

\section{Introduction}\label{sec:introduction}

As a result of breakthroughs in biotechnology and high-throughput experiments thousands of regulatory and protein-protein interactions have been revealed and genome-wide protein-protein interaction (PPI) data are now available. Protein-protein interactions are one of the most important components of biological networks as they are fundamental to the functioning of cells. To gain a better understanding of why these interactions take place, it is necessary to view them from an evolutionary perspective. The evolutionary history of PPI networks can help answer many questions about how present-day networks have evolved and provide valuable insight into molecular mechanisms of network growth \cite{Pere07,Kreim08}. However, inferring network evolution history is a statistical and computational challenging problem as PPI networks of extant organisms provide only snapshots in time of the network evolution. There has been recent work on reconstructing ancestral interactions (e.g. \cite{Dut07,Gibs09,Patro12}). 
The main growth mechanism of PPI network is gene duplication and divergence (mutations) \cite{Wag01}:  all proteins in a family evolve from a common ancestor through gene duplications and mutations  and the protein network reflects the entire history of the genome evolution \cite{Vazquez_2003}. In this paper we follow \cite{Li_2013} 
and we develop computational methods to infer the growth history and the parameters under the given model 
incorporating not only the topology of observed networks, but also
the duplication history of the proteins contained in the networks.
In their article \cite{Li_2013} propose a maximum likelihood
approach. The authors establish a neat representation of the likelihood function and it is this representation which is used in this article. 
The duplication history of the proteins can be inferred independently by phylogenetic analysis \cite{Patro12,Pin07}.

The approach we adopt here is first
to obtain a numerically stable estimate of the likelihood function, under fixed parameters; this achieved via the sequential Monte Carlo (SMC) method (see \cite{Doucet_2000} and \cite{Gordon_1993}). This approach can then be used to infer the parameters of the model, from a Bayesian perspective, as well as the growth history, via a Markov chain Monte Carlo
(MCMC) method. To the best of our knowledge, this has not been considered in the literature, although related ideas have appeared for simpler models in \cite{wang}.
Our computational strategy not only improves on likelihood estimation in comparison to \cite{Li_2013}, but also provides a natural setup to perform posterior inference on the parameters of interest.

This article is structured as follows. In Section \ref{sec:modelsandmethods} we detail the model and associated computational method for statistical inference. In Section \ref{sec:results}
our numerical results are presented. In Section \ref{sec:discussion} the article is concluded with some discussion of future work.

\section{Model and Methods}\label{sec:modelsandmethods}
We follow similar notation and exposition as in \cite{Li_2013} to introduce the protein-protein interaction network, its duplication history, and the duplication-mutation with complementarity  (DMC) model \cite{Vazquez_2003}. In particular, the notions of adjacency and duplication are made concrete there.
We also introduce the associated Bayesian inference problem with which this work is primarily concerned (i.e., that of inferring the parameters of the DMC model). We then describe a particle marginal Metropolis-Hastings (PMMH) algorithm \cite{Andrieu_2010} that can be used to perform such inference.

\subsection{PPI network and DMC model}
Consider an undirected graph $G$ without multiple edges and self-loops, where the nodes represent proteins and the edges represent interactions between those proteins. Such a graph is called a PPI network, and as in \cite{Li_2013}, we denote the vertex set by $V(G)$, the edge set by $E(G)$, and the number of nodes in $G$ by $|V(G)|$. All nodes which are adjacent to a node $v$ (not including $v$ itself) comprise the neighbourhood of $v$, and that neighbourhood is denoted by $N_G(v)$.

We assume that $G$ evolved from a seed graph $G_0$ via a series of duplication, mutation, and homodimerization steps under a DMC model. Under the DMC model, at each time step $t$, the graph $G_t$ evolves from $G_{t-1}$ by the following processes in order:
\begin{enumerate}
 \item{The anchor node $u_t$ is chosen uniformly at random from $V(G_{t-1})$, and a duplicate node $v_t$ is added to $G_{t-1}$ and connected to every member of $N_{G_{t-1}}(u_t)$. This is the duplication step, and it yields an intermediary graph denoted $G_{t-1}^*$.}
 \item{For each $w \in N_{G_{t-1}^*}(u_t)$, we uniformly choose one of the two edges in $\{(u_t,w),(v_t,w)\} \subseteq E(G_{t-1}^*)$ at random and delete it with probability $(1-p)$. This is the mutation step, and the parameter $p$ is henceforth referred to as the mutation parameter.}
 \item{The anchor node $u_t$ and the duplicate node $v_t$ are connected with probability $p_c$ to finally obtain $G_t$. This is the homodimerization step, and the parameter $p_c$ is henceforth known as the homodimerization parameter.}
\end{enumerate}

The DMC model is Markovian, and we denote the transition density at time $t$ (which encompasses the three afore-mentioned steps) by $p_{\mathcal{M}}(G_t \mid G_{t-1})$, where $\mathcal{M} := (p, p_c)$. If we assign to a seed graph some prior density $p_{\mathcal{M}}(G_0)$, then the density of the observed graph $G$ will be
\begin{align}\label{eq:densityG1}
 p_{\mathcal{M}}(G) = \sum_{\mathcal{H} \backslash \{G_n\}} \bigg[ p_{\mathcal{M}}(G_0) \prod_{t=1}^{n}p_{\mathcal{M}}(G_t \mid G_{t-1}) \bigg],
\end{align}
where $G = G_n$, $n = |V(G)| - |V(G_0)|$ and $\mathcal{H}=(G_0,G_1,\dots,G_n=G)$ 
denotes the collection of  growth histories.
In this work, a seed graph will always be the graph consisting of two connected nodes; thus, $|V(G_0)| = 2$ and $p_{\mathcal{M}}(G_0) = 1$. Note that we are summing over all possible growth histories by which $G$ can evolve from a seed graph. Also note that a growth history $\mathcal{H}$ induces a unique sequence of duplicate nodes, $\theta(\mathcal{H})=(v_1,\dots,v_n)$ \cite{Li_2013}.

\subsection{Bayesian inference}
In practice, one will not have access to the parameters $(p, p_c)$ and they must be inferred given $G$. Thus, in the Bayesian setting, our objective is to consider the posterior density
\begin{align}\label{eq:densityG2}
 \pi(\mathcal{M} \mid G) \propto p(\mathcal{M})p_{\mathcal{M}}(G),
\end{align}
where $p(\mathcal{M})$ is some proper prior for $(p, p_c)$ that we assume can easily be computed (at least pointwise up to a normalising constant).

The total number of growth histories grows exponentially with $n$ \cite{Li_2013}, and so any computations involving \eqref{eq:densityG1}, and thus \eqref{eq:densityG2}, 
(e.g.~the evaluation of $p_{\mathcal{M}}(G)$)
could potentially become very expensive. In the following sections, we reformulate the inference problem in the same manner as in \cite{Li_2013} to alleviate this issue.

\subsection{Duplication history}
As in \cite{Li_2013}, let  $\Gamma$ be a binary forest, i.e.~a collection of rooted binary trees. The authors of \cite{Li_2013} describe a scheme that encodes the duplication history of a growth history $\mathcal{H}$ within a series of duplication forests, $(\Gamma_0,\Gamma_1,\dots,\Gamma_n)$, where each forest $\Gamma_t$ corresponds to a graph $G_t$. We describe that scheme here.

Consider a trivial forest $\Gamma_0$, whose only two isolated trees each consist of a single node. Each of those isolated nodes will correspond to a node within the seed graph $G_0$. To build $\Gamma_1$ from $\Gamma_0$, one replaces an anchor node $u_1$ from $\Gamma_0$ with a subtree, $\{u_1,v_1\}$, consisting of two leaves ($v_1$ is the duplicate node included $G_1$ but not $G_0$). This process continues until one builds the series of forests $(\Gamma_0,\Gamma_1,\dots,\Gamma_n = \Gamma)$ to correspond to $\mathcal{H}$.

As highlighted in \cite{Li_2013}, the duplication forest $\Gamma$ (corresponding to $G$) is uniquely determined by $\mathcal{H}$ and a list of anchor nodes, $\pi=(u_1,\dots,u_n)$. 
The important thing to emphasize here, is that given the duplication 
forest $\Gamma$ and $G$, one now only needs to infer the duplication nodes sequence 
$\theta(\mathcal{H})=(v_1,v_2,\ldots, v_n)$ to reconstruct the complete growth history $\mathcal{H}$. For instance, at the first step backwards, knowledge of $\Gamma_n=\Gamma$ 
together with $v_n$ uniquely identifies the anchor node $u_n$, thus one can reconstruct
$G_{n-1}$ and $\Gamma_{n-1}$; this is then repeated for the remaining backward steps.
Thus, given $\theta(\mathcal{H})$, 
 one can construct the growth history $\mathcal{H}$ backward-in-time using the backward operators defined in \cite[Section 2.4]{Li_2013}, which construct $G_{t-1}, \Gamma_{t-1}$  deterministically given $(G_{t}, \Gamma_{t}, v_t)$, for $t=n,n-1,\ldots,1$.

\begin{figure}[!h]
\begin{centering}\quad \quad 
\includegraphics[scale=0.4]{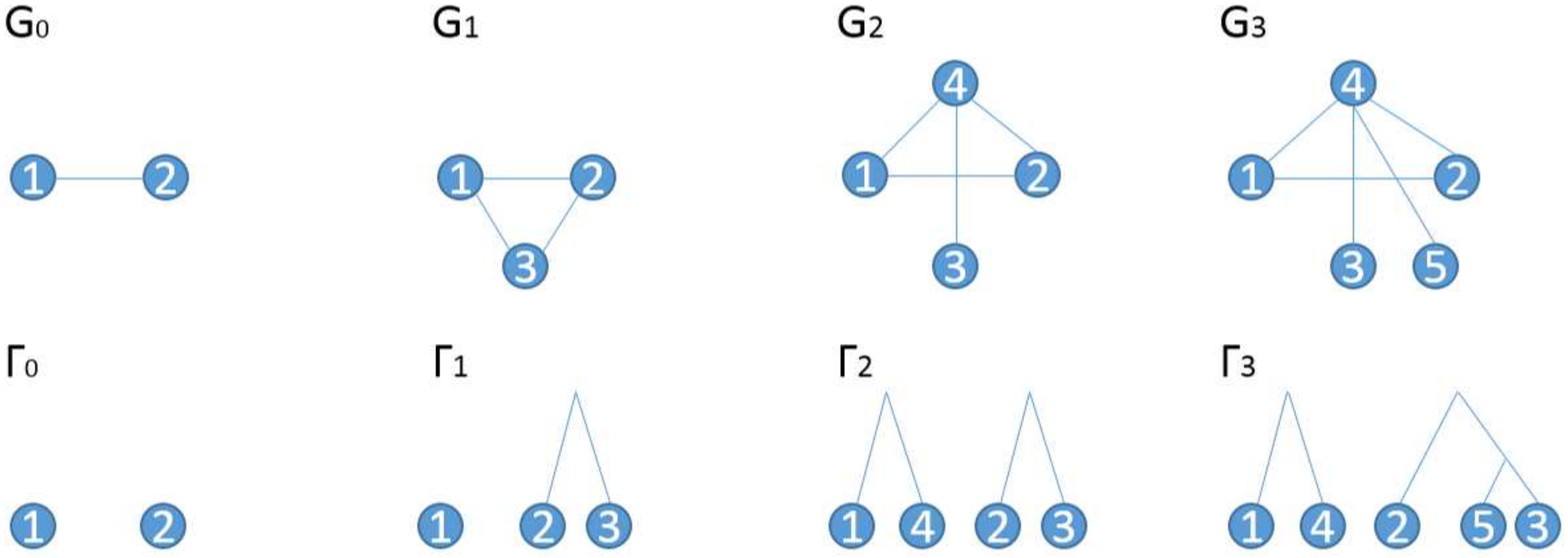}
\end{centering}
\vspace{-2.1cm}
\caption{An example growth history for a network together with the corresponding 
history of the duplication forest. In this example, $(u_1, u_2, u_3)= (2,1,3)$ and
$(v_1, v_2, v_3)=(3,4,5)$.}
\label{fig:model}
\end{figure}

\subsection{Bayesian inference given the duplication history}

Now suppose that in addition to $G$, a practitioner is given $\Gamma$ corresponding to $G$. Our new objective -- and the primary inference problem with which this work is concerned -- is to consider the posterior density $\pi(\mathcal{M}|G,\Gamma)$. 
Notice that we have the joint distribution:
\begin{equation*}
\pi(\mathcal{M},\{G,\Gamma\},\theta) = p(\mathcal{M})\,p_{\mathcal{M}}(G_0^{\theta},\Gamma_0^{\theta}) \prod_{t=1}^{n}p_{\mathcal{M}}(G_t^{\theta}, \Gamma_{t}^{\theta} \mid G_{t-1}^{\theta}, \Gamma_{t-1}^{\theta})
\end{equation*}
where $\theta=(v_1,\ldots,v_n)$ is a sequence of duplication nodes compatible with 
the observed $G,\Gamma$, and $G_0^{\theta},\Gamma_0^{\theta},\ldots, G_n^{\theta},\Gamma_n^{\theta}$
the corresponding reconstructed   history. We are thus interested in the 
parameter posterior:
\begin{align}\label{eq:densityG3}
\pi(\mathcal{M} \mid G, \Gamma ) &\propto p(\mathcal{M})p_{\mathcal{M}}(G, \Gamma), \\
p_{\mathcal{M}}(G, \Gamma) &= \sum_{{\theta}|G,\Gamma} \bigg[ p_{\mathcal{M}}(G_0^{\theta}, \Gamma_{0}^{\theta}) \prod_{t=1}^{n}p_{\mathcal{M}}(G_t^{\theta}, \Gamma_{t}^{\theta} \mid G_{t-1}^{\theta}, \Gamma_{t-1}^{\theta}) \bigg], \nonumber
\end{align}
The density $p_{\mathcal{M}}(G_0, \Gamma_{0})$ is typically a trivial term which could be ignored in practice. As the duplication forest $\Gamma$ limits the number of allowable anchor-and-duplicate node pairs, one can see that the number of possible growth histories is reduced.

\subsection{Methods}\label{sec:methods}
We will now present an SMC algorithm that can sample the latent growth histories from the DMC model given the fixed parameters $\mathcal{M}:=(p, p_c)$. We then show that this algorithm can be employed within a PMMH algorithm, as in \cite{Andrieu_2010}, to sample from the posterior \eqref{eq:densityG3} and infer $\mathcal{M}$ (and even $\mathcal{\theta} | G,\Gamma$).

An SMC algorithm simulates a collection of $N$ samples (or, particles) sequentially along the index $t$ via importance sampling and resampling techniques to approximate a sequence of probability distributions of increasing state-space, which are known pointwise up to their normalising constants. 
In this work, we use the SMC methodology to sample from the posterior distribution 
of the latent duplication history: 
\begin{align*}
 p_{\mathcal{M}}(\theta \mid G, \Gamma) \propto p_{\mathcal{M}}(G_0^{\theta},
  \Gamma_{0}^{\theta}) \prod_{t=1}^{n}p_{\mathcal{M}}(G_t^{\theta}, \Gamma_{t}^{\theta} \mid G_{t-1}^{\theta}, \Gamma_{t-1}^{\theta})
\end{align*}
backwards along the index $t$ via Algorithm \ref{alg:SMC} in the Appendix. The technique provides an unbiased estimate of the normalising constant \cite[Theorem 7.4.2]{Delmoral_2004}, $p_{\mathcal{M}}(G, \Gamma)$:
\begin{equation}\label{eq:normalizingconstantest}
 \hat{p}_{\mathcal{M}}(G, \Gamma) = \prod_{t=0}^{n-1} \bigg[ \frac{1}{N} \sum_{i=1}^N W_t^i\bigg],
\end{equation}
where each $W_t^i$ is an un-normalised importance weight computed in Algorithm \ref{alg:SMC}. Note that under assumptions on the model, if $N>cn$ for some $c<\infty$, then the relative variance of the estimate is $\mathcal{O}(n/N)$; see \cite{cerou1}. It is remarked that, as in \cite{wang}, one could also use the discrete particle filter \cite{fearnhead}, with a possible improvement over the SMC method detailed in 
Algorithm~\ref{alg:SMC}; see \cite{wang} for some details.

This SMC can be employed within a PMMH algorithm to target the posterior of $\mathcal{M}$ in \eqref{eq:densityG3}. One can think of the deduced method as an 
MCMC algorithm running on the marginal $\mathcal{M}$-space, 
but with the SMC unbiased estimate $\hat{p}_{\mathcal{M}}(G,\Gamma)$ replacing the unknown 
likelihood $\hat{p}_{\mathcal{M}}(G,\Gamma)$. 
More analytically, we can consider all random variables involved in the method
and write down the equilibrium distribution in the enlarged state space,
with $\mathcal{M}$-marginal the target posterior $p_{\mathcal{M}}(G,\Gamma)$.
Following \cite{Andrieu_2010} and letting $\phi_t^i$ denote a sample $(G_t,\Gamma_t)$  at time $t$,
the extended equilibrium distribution is written as:
\begin{align}
 \pi^N &\left( l, \mathcal{M}, a_{1:n-1}^{1:N}, \phi_{0:n-1}^{1:N} \mid G, \Gamma \right) = 
\nonumber 
 \\
& 
 \frac{\pi\left(\mathcal{M},\phi_{0:n-1}^{l}\,|\,G,\Gamma\right)}{N^{n}} \cdot
 \frac{\Psi_{\mathcal{M}}(a_{1:n-1}^{1:N},\phi_{0:n-1}^{1:N})}{q_{\mathcal{M}}(v_n^{a_{n-1}^l}) 
 \left(\prod_{t=1}^{n-1} {w}_{t}^{a_{t}^l} q_{\mathcal{M}}(v_t^{a_{t-1}^l})\right)}\ , 
 \label{eq:extendedtarget}
\end{align}
where $\Psi_{\mathcal{M}}$ is the probability of all the variables associated to Algorithm \ref{alg:SMC}, with $a_k^j,l\in\{1,\dots,N\}$ and the $\phi$'s being the simulated variables at each step of 
Algorithm \ref{alg:SMC}. 


A PMMH algorithm (see Algorithm \ref{alg:PMMH}) samples from \eqref{eq:extendedtarget}, and
one can remove the auxiliary variables from the samples to obtain draws 
for the parameters from \eqref{eq:densityG3}. Furthermore, one could even save the sampled growth histories with particle index $l$ to obtain draws from the joint posterior $\pi(\mathcal{M},\theta \mid G, \Gamma )$.
However, in this work, we are primarily interested in the inference of $\mathcal{M}$.

\section{Results}\label{sec:results}
The variance of the estimate \eqref{eq:normalizingconstantest} plays a crucial role in the performance of Algorithm \ref{alg:PMMH}, as \eqref{eq:normalizingconstantest} is used to compute the acceptance probability within the PMMH algorithm. Thus, we first tested the variability of \eqref{eq:normalizingconstantest} as computed by Algorithm \ref{alg:SMC} to understand how the variance changes with $|V(G)|$. We then ran Algorithm \ref{alg:PMMH} to sample from the posterior \eqref{eq:densityG3} and infer $\mathcal{M}$ for a given 
pair of observations  $(G, \Gamma)$. We present the details of those experiments below.

\subsection{Variance of $\hat{p}_{\mathcal{M}}(G, \Gamma)$}
We simulated a graph $G$ and a forest $\Gamma$ from the DMC model \cite{Vazquez_2003} with the parameters set as $(p = 0.7, p_c = 0.7)$, where $|V(G)| = 40$. We saved each pair $(G_t, \Gamma_t)$ for $1 \leq t \leq 40$, and we ran Algorithm \ref{alg:SMC} fifty times per pair (with $N=|V(G_t)|*5$) to compute fifty unbiased estimates of $p_{\mathcal{M}}(G_t, \Gamma_t)$ for $1 \leq t \leq 40$. In the top of Figure \ref{fig:smctest} in Section \ref{sec:figures}, we plot the relative variance of the estimate (or, the variance divided by the square of the expected value) per each value of $|V(G_t)|$. We repeated the experiment two more times, with $N=|V(G_t)|*10$ and $N=|V(G_t)|*20$, and the associated output is also presented in Figure \ref{fig:smctest}.

As remarked above,  if $N>cn$ for some $c<\infty$, then the relative variance of $\hat{p}_{\mathcal{M}}(G_t, \Gamma_t)$ is $\mathcal{O}(n/N)$.
Figure \ref{fig:smctest} confirms that the variance increases linearly and that increasing the value of $N$ with $|V(G)|$ (at least linearly) will help to control the variance. However, the plots show that the relative variance is still high, which means that $N$ will have to be large to ensure satisfactory performance of the PMMH in practice.

\subsection{Parameter inference}
We separately simulated a graph $G$ and a forest $\Gamma$ from the DMC model with the parameters again set as $(p = 0.7, p_c = 0.7)$, and we set $|V(G)| = 15$. Given only $(G, \Gamma)$, we inferred $(p, p_c)$ with each parameter having a uniform prior on the interval $[0.1,0.9]$. We set the number of particles within Algorithm \ref{alg:SMC} to be $N=2000$, and we ran the PMMH algorithm to obtain 10,000 samples from the extended target.

Figure \ref{fig:simdata} in Section \ref{sec:figures} illustrates good mixing of the PMMH algorithm and accurate inference of the parameters $(p, p_c)$. The trace plots show that the algorithm is not sticky, and the autocorrelation functions give evidence to an approximate independence between samples. The posterior densities are also interesting, in that they are clearly different from the uniform priors and they show that the PMMH algorithm spends a majority of the computational time sampling the true parameter values.

\section{Discussion}\label{sec:discussion}
We have introduced a Bayesian inferential framework for the DMC model \cite{Vazquez_2003}, 
where, as in \cite{Li_2013}, one assumes the pair $(G, \Gamma)$ is observed and the parameters $(p, p_c)$ are unknown. We then described how an SMC algorithm can be used to simulate growth histories and ultimately be employed within PMMH to target the posterior distribution of the parameters \eqref{eq:densityG3}, thereby opening up the possibility of performing Bayesian inference on the DMC model.
 
 Numerical tests demonstrated that Algorithm \ref{alg:SMC} can have a high variability when $|V(G)|$ is large and $N$ is not sufficiently high, and this limits the scope of the inference problem which can be tackled using the complete Algorithm \ref{alg:PMMH}. However, the proposals used in the example experiments within Algorithms \ref{alg:SMC} and \ref{alg:PMMH} are naive, 
 as the method simply chooses a candidate duplicate node $v_t^{i}$ at random
 from all permitted nodes given the current  $G_t^{i}, \Gamma_t^{i}$. 
 It is reasonable to assume that more sophisticated proposal densities could reduce the variance of the SMC and/or improve the mixing of the PMMH, thereby allowing one to perform inference when $|V(G_t)|$ is large and $N$ is smaller. This could be explored in a future work.

\section*{Acknowledgements}
This research was funded by the EPSRC grant ``Advanced Stochastic Computation for Inference from Tree, Graph and Network Models'' (Ref: EP/K01501X/1). AJ was additionally 
supported by a Singapore Ministry of Education Academic Research Fund Tier 1 grant (R-155-000-156-112) and is also affiliated with the risk management institute and the centre for quantitative finance at the National University of Singapore.

\appendix

\section{Figures}\label{sec:figures}
\begin{figure}[H]
\begin{center}
\textbf{Variability of $\hat{p}_{\mathcal{M}}(G, \Gamma)$} \\
\scalebox{0.4}{\includegraphics[trim = 15mm 10mm 13mm 10mm, clip]{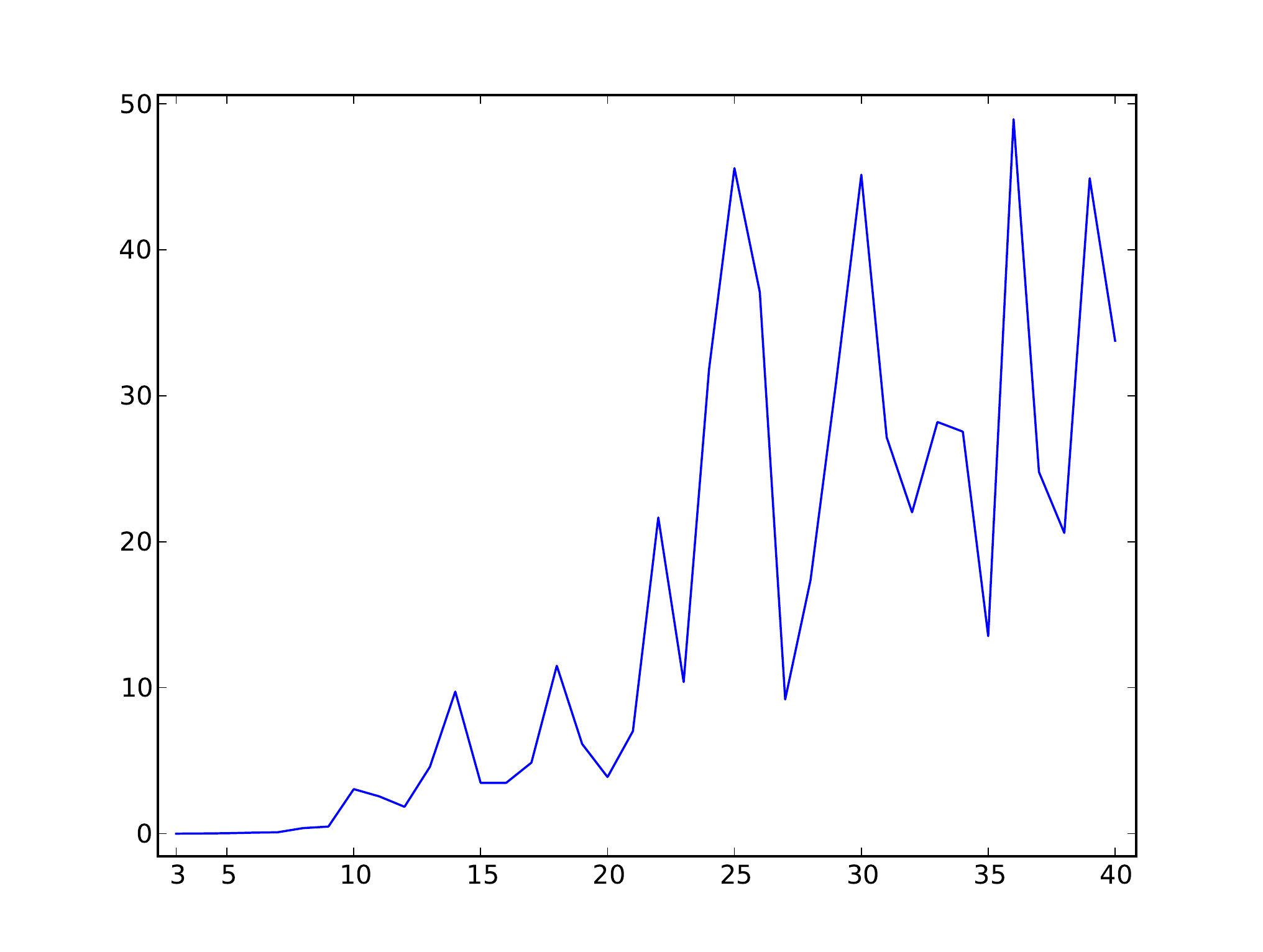}}\\
\scalebox{0.4}{\includegraphics[trim = 15mm 10mm 13mm 10mm, clip]{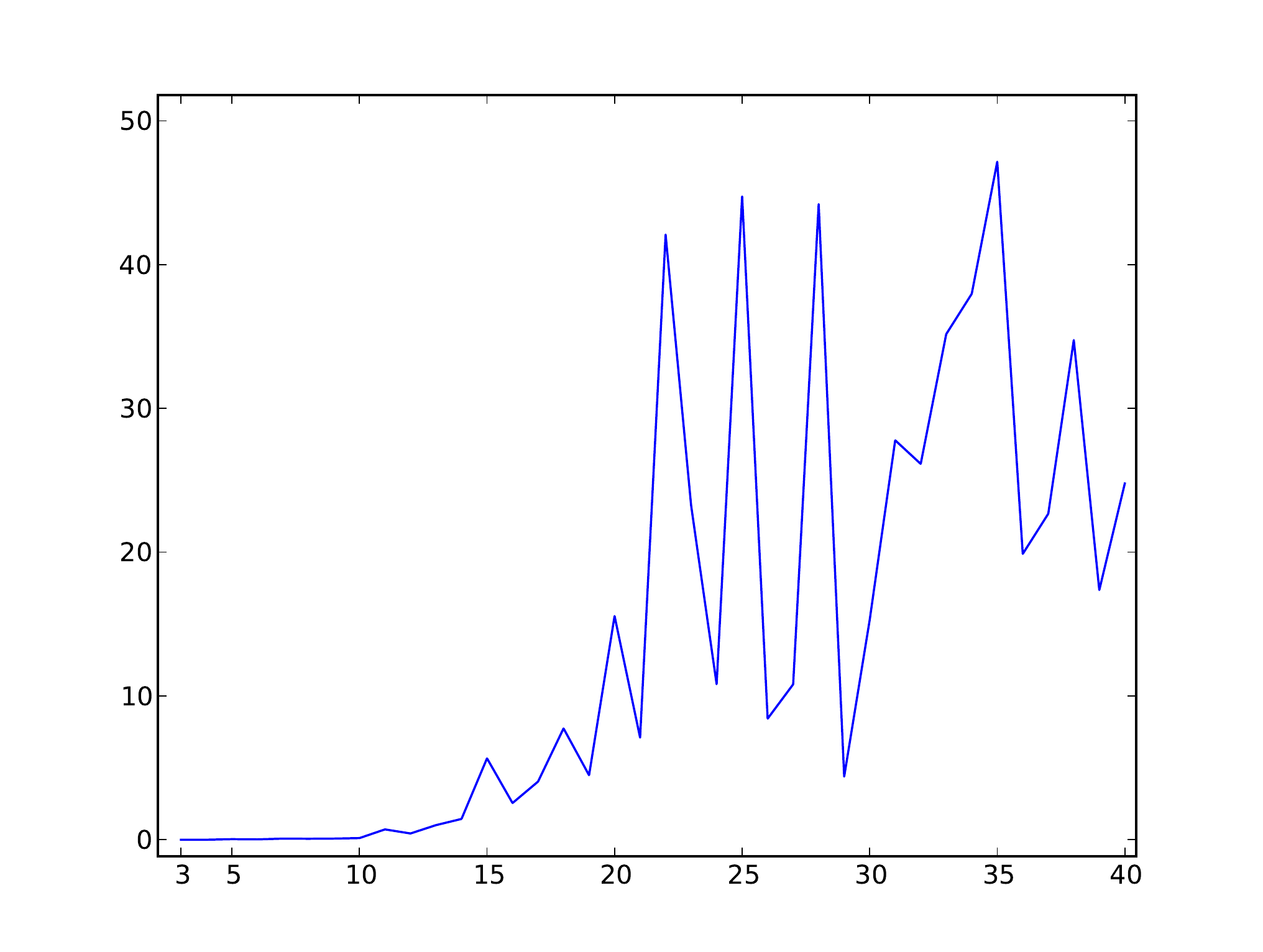}}\\
\scalebox{0.4}{\includegraphics[trim = 15mm 10mm 13mm 10mm, clip]{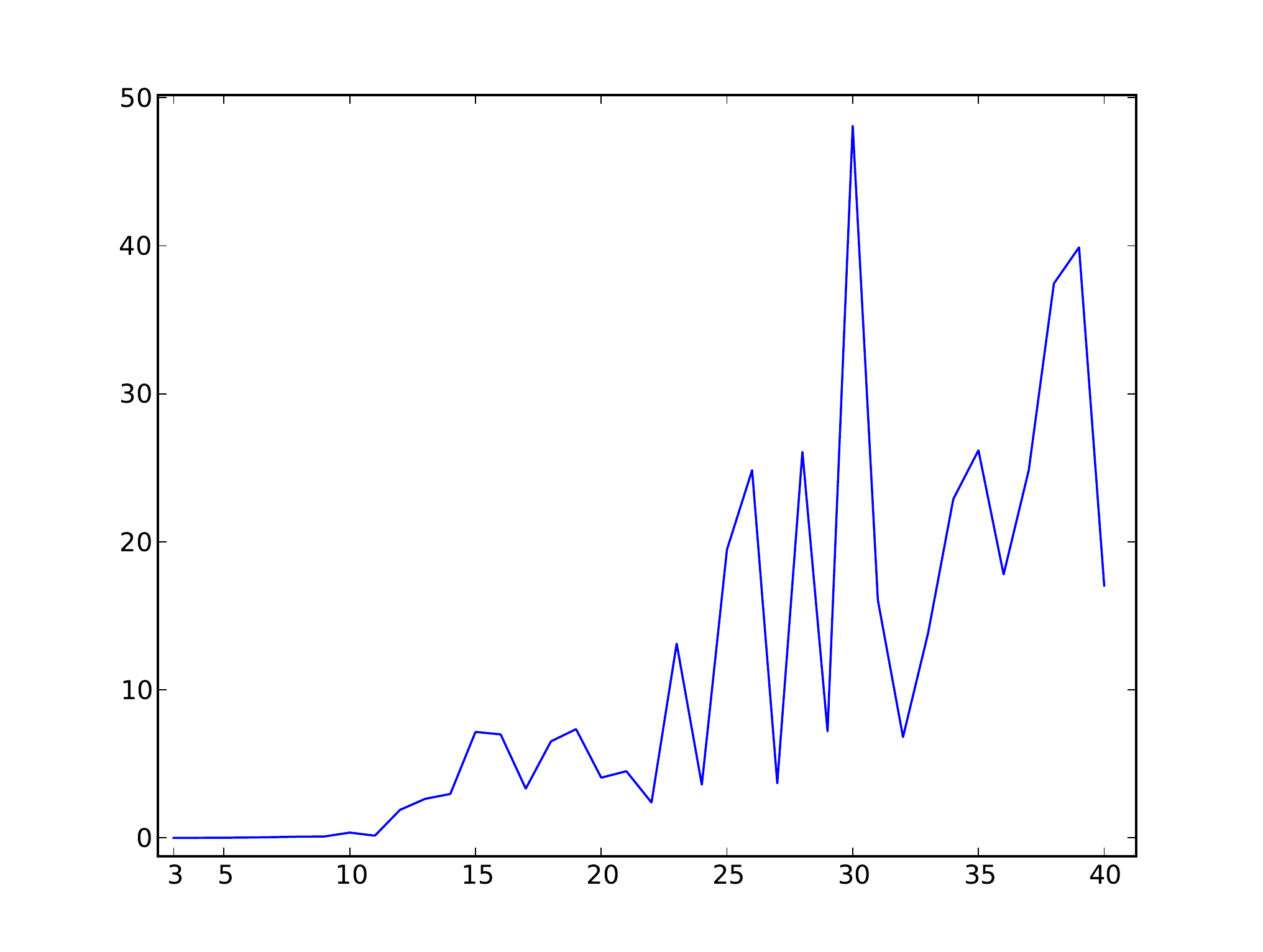}}
\vspace{-0.3cm}
\caption{All plots illustrate the relative variance of $\hat{p}_{\mathcal{M}}(G_t, \Gamma_t)$, per $|V(G_t)|$ on the horizontal axis; the relative variance is the variance divided by the square of the expected value. In the top plot, the number of SMC particles used to compute each $\hat{p}_{\mathcal{M}}(G_t, \Gamma_t)$ is $|V(G_t)|*5$. In the middle and bottom plots, that number is $|V(G_t)|*10$ and $|V(G_t)|*20$, respectively. Recall that the seed graph, $G_0$, has two nodes, and note that we did not compute $\hat{p}_{\mathcal{M}}(G_0, \Gamma_0)$ because it is trivial.}
\label{fig:smctest}
\end{center}
\end{figure}

\begin{figure}[H]
\begin{center}
\textbf{Parameter inference} \\
\scalebox{0.35}{\includegraphics[trim = 15mm 10mm 13mm 10mm, clip]{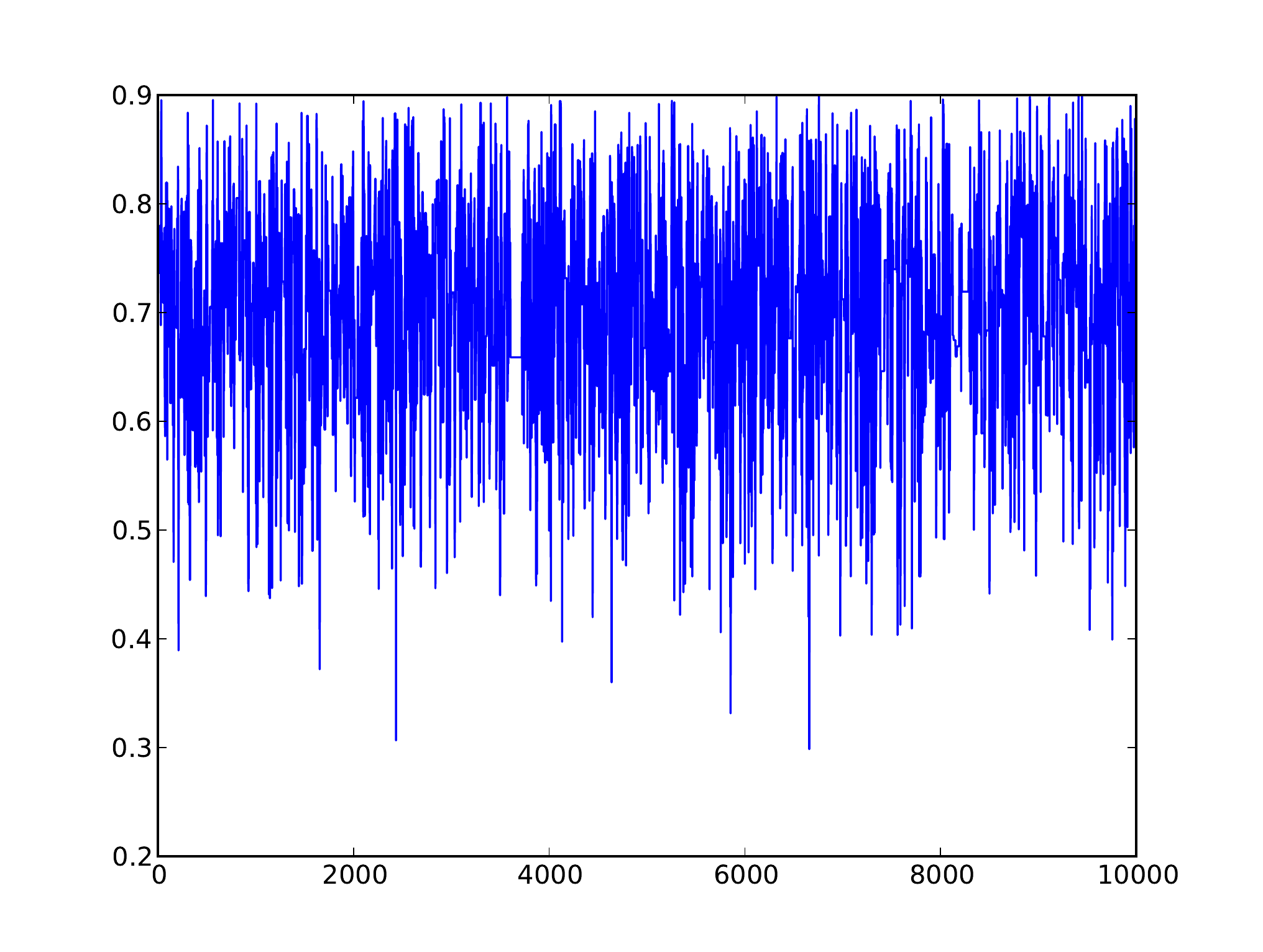}}
\scalebox{0.35}{\includegraphics[trim = 15mm 10mm 13mm 10mm, clip]{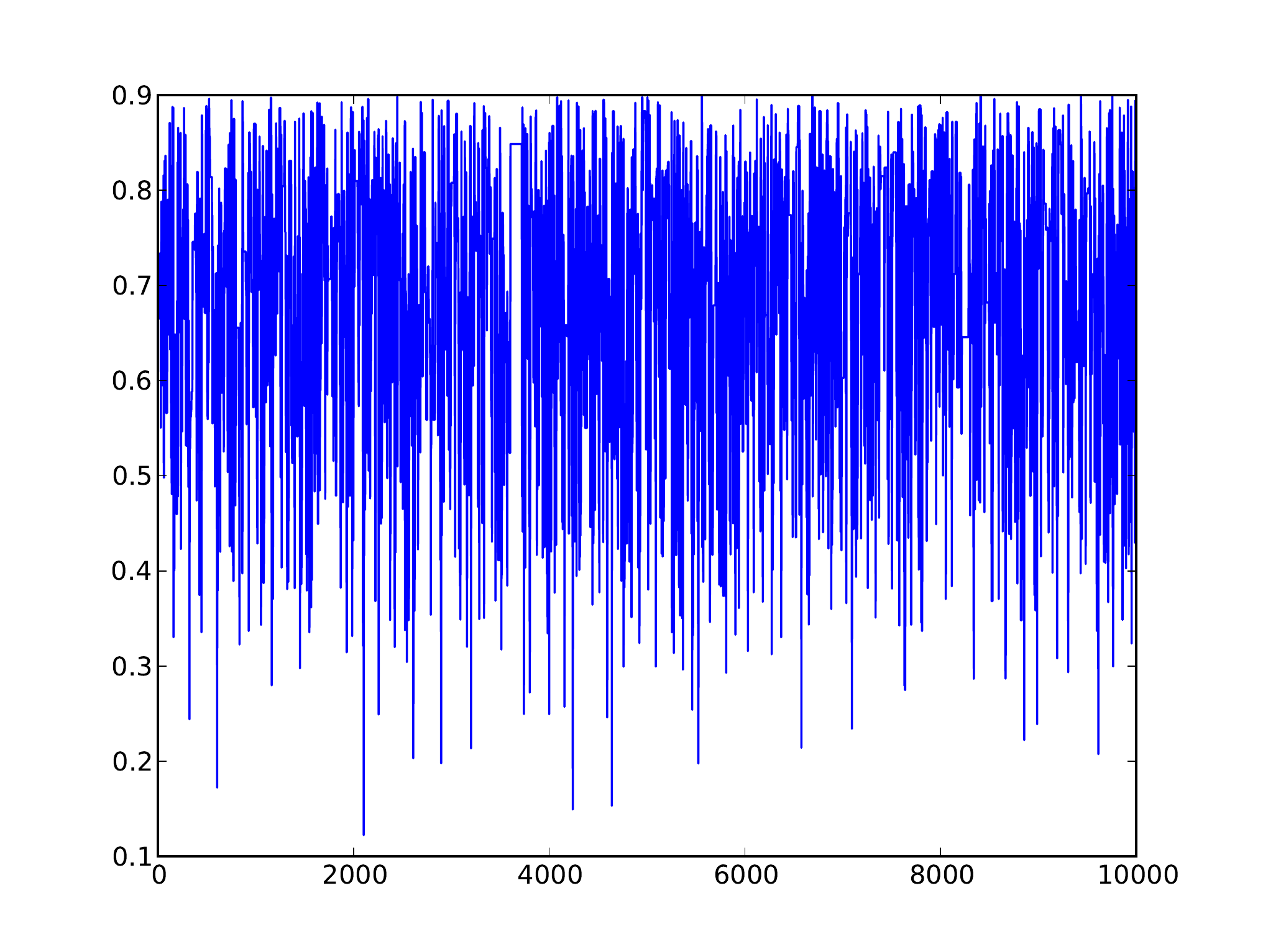}}\\
\scalebox{0.35}{\includegraphics[trim = 15mm 10mm 13mm 10mm, clip]{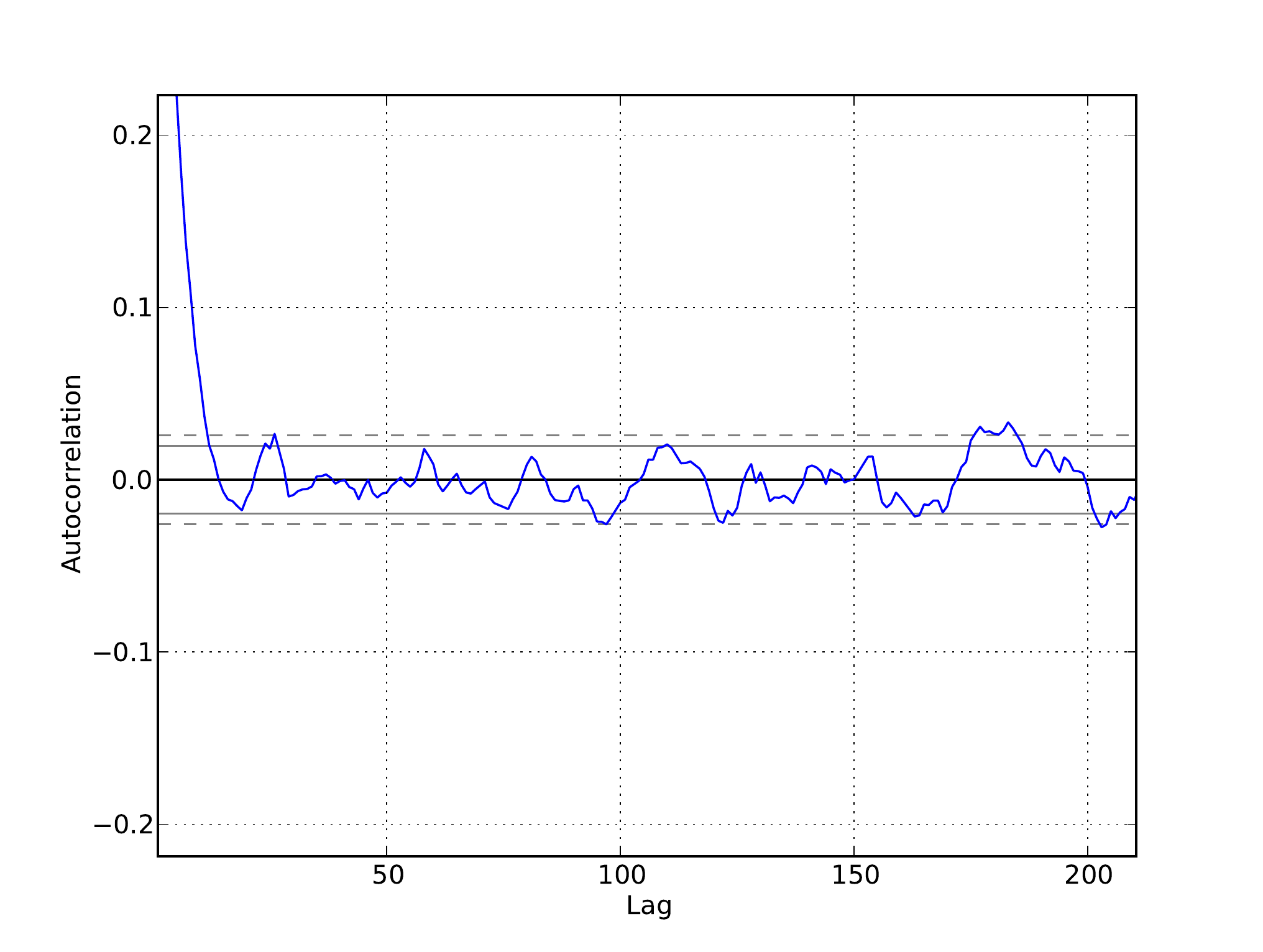}}
\scalebox{0.35}{\includegraphics[trim = 15mm 10mm 13mm 10mm, clip]{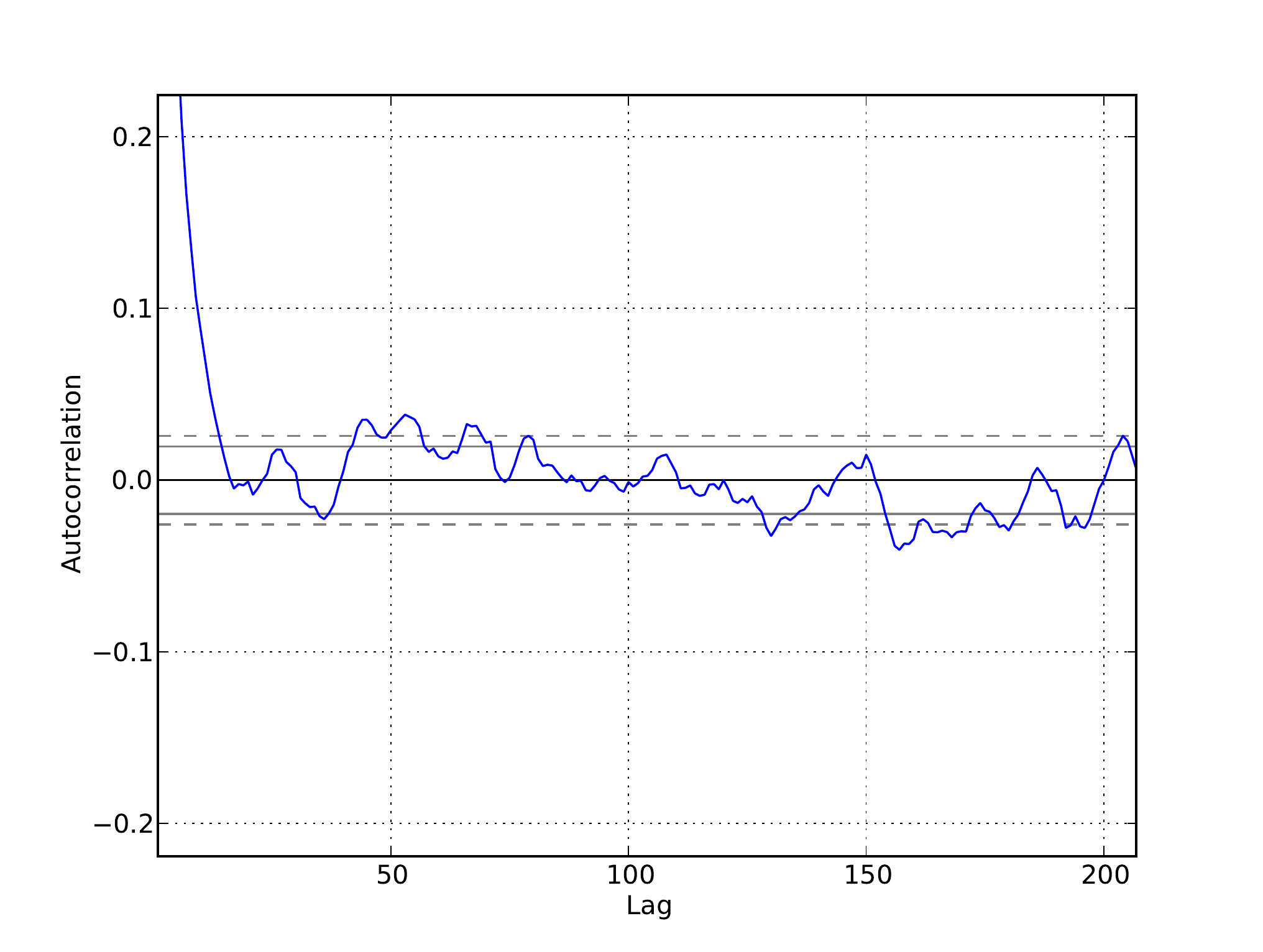}}\\
\scalebox{0.35}{\includegraphics[trim = 17mm 10mm 13mm 10mm, clip]{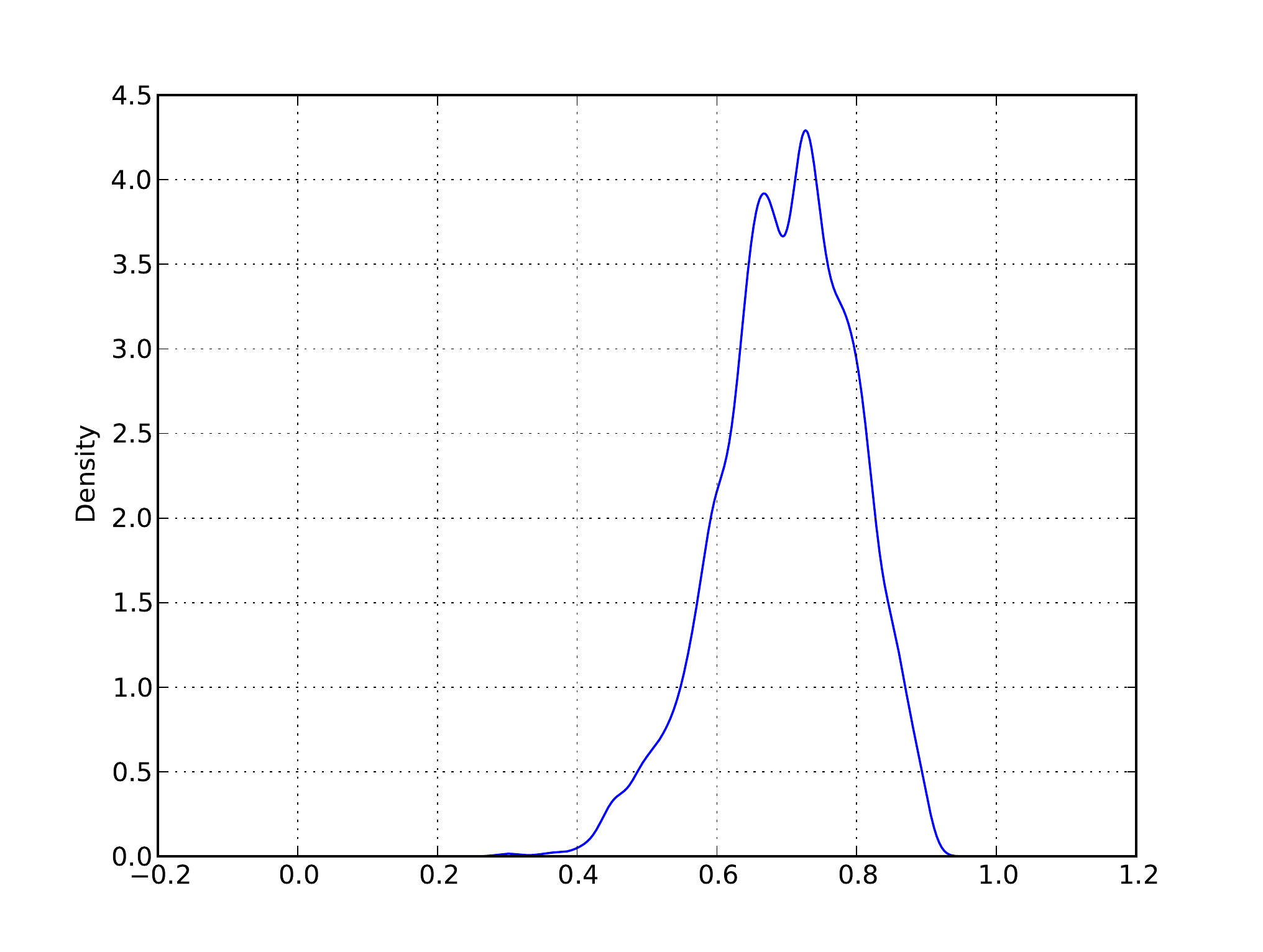}}
\scalebox{0.35}{\includegraphics[trim = 17mm 10mm 13mm 10mm, clip]{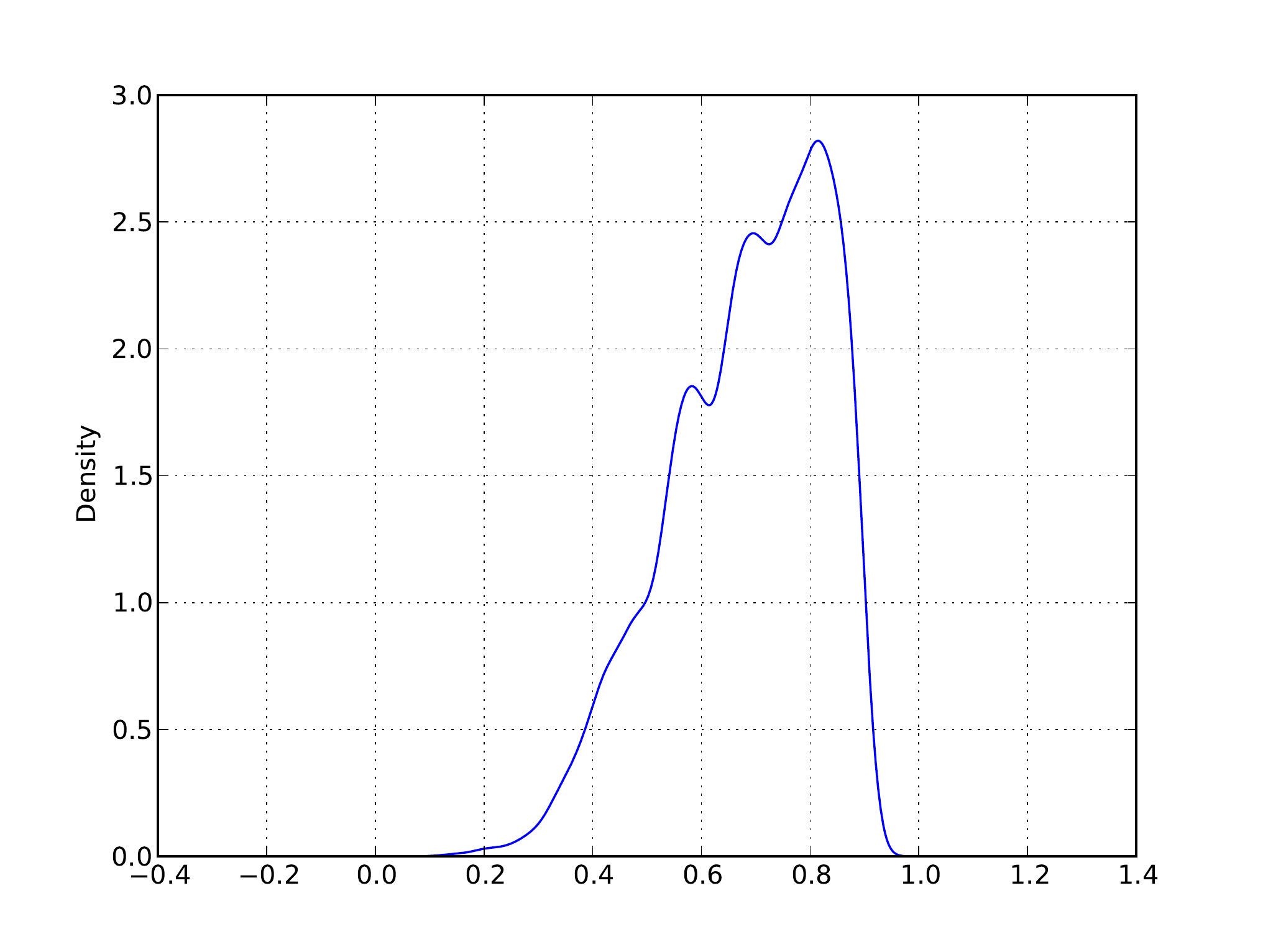}}
\vspace{-0.3cm}
\caption{Plots associated with $p$ ($p_c$) are at left (right). The top figures are trace plots, with PMMH iteration running along the horizontal axes and parameter value running along the verticals. The middle figures are plots of the autocorrelation functions (with lag running along the horizontal axes), and at the bottom we present the parameter posterior densities.}
\label{fig:simdata}
\end{center}
\end{figure}

\section{Algorithm Summaries}\label{sec:algosums}
\begin{algorithm}[H]
\begin{itemize}
\item{ Step 0: Input an observed graph $G = G_n$ and a corresponding observed forest $\Gamma = \Gamma_n$, where $G$ is not a seed graph.}
\item{ Step 1: Set $t = n$. For $i\in\{1,\dots,N\}$, sample a subtree with two nodes uniformly at random from $\Gamma_t^{i}$, and chose one of the two nodes uniformly as the proposed 
duplicate node $v_t^{i}$ (thus the other will be 
the anchor node). 
Using the backward operators defined in \cite[Section 2.4]{Li_2013}, construct each $(G_{t-1}^i, \Gamma_{t-1}^i)$ from the subtrees and $(G_{t}^{i}, \Gamma_{t}^{i})$. For $i\in\{1,\dots,N\}$, compute the un-normalised weight
\begin{equation*}
{W}_{t-1}^i = \frac{p_{\mathcal{M}}(G_t^{i}, \Gamma_{t}^{i} \mid G_{t-1}^i, \Gamma_{t-1}^i)}{q_{\mathcal{M}}(v_t^i)},
\end{equation*}
where $q_{\mathcal{M}}$ is the density of the proposal mechanism used to sample $\{u_t^i\}$.}
\item{ Step 2: If $\{G_{t-1}^{1:N}\}$ are not seed graphs, then set $t=t-1$ and continue to Step 3. Otherwise, the algorithm terminates.}
\item{ Step 3: For $i\in\{1,\dots,N\}$, sample $a_{t}^i\in\{1,\dots,N\}$ from a discrete distribution on $\{1,\dots,N\}$ with $j^\text{th}$ probability $w_{t}^j \propto W_{t}^j$.  The sample $\{ a_{t}^{1:N} \}$ are the indices of the resampled particles. Set all normalised weights equal to $N^{-1}$.}
\item{ Step 4: For $i\in\{1,\dots,N\}$, sample a subtree 
with two nodes uniformly at random  from the resampled forest $\Gamma_t^{a_t^i}$,
and select uniformly one of the two nodes 
as the proposed  duplicate node $v_t^{i}$. 
Construct $(G_{t-1}^i, \Gamma_{t-1}^i)$ from $v_t^{i}$, $(G_{t}^{a_t^i}, \Gamma_{t}^{a_t^i})$.
For $i\in\{1,\dots,N\}$, compute the un-normalised weight
\begin{equation*}
{W}_{t-1}^i = \frac{p_{\mathcal{M}}(G_{t}^{a_t^i}, \Gamma_{t}^{a_t^i} \mid G_{t-1}^i, \Gamma_{t-1}^i)}{q_{\mathcal{M}}(v_t^i)}\ .
\end{equation*}
Return to Step 2.}
\end{itemize}
\caption{\label{alg:SMC}Sequential Monte Carlo (SMC)}
\end{algorithm}

\begin{algorithm}[H]
\begin{itemize}
 \item{ Step 0: Set $r=0$. Sample $\mathcal{M}^{(r)}\sim p(\cdot)$. All remaining random variables can be sampled from their full conditionals defined by the target \eqref{eq:extendedtarget}:
 
 - Sample $\phi_{0:n-1}^{(r),1:N}, a_{1:n-1}^{(r),1:N} \sim \Psi_{\mathcal{M}^{(r)}}(\cdot)$ via Algorithm \ref{alg:SMC}.

 - Choose a particle index $l^{(r)} \propto W_0^{(r),l^{(r)}}$.
 
 Finally, calculate $\hat{p}_{\mathcal{M}^{(r)}}(G, \Gamma)$ via \eqref{eq:normalizingconstantest}.}
 
 \item{ Step 1: Set $r=r+1$. Sample $\mathcal{M}^{*}\sim q(\cdot|\mathcal{M})$. All remaining random variables can be sampled from their full conditionals defined by the target \eqref{eq:extendedtarget}:
 
 - Sample $\phi_{0:n-1}^{*,1:N}, a_{1:n-1}^{*,1:N} \sim \Psi_{\mathcal{M}^{*}}(\cdot)$ via Algorithm \ref{alg:SMC}.

 - Choose a particle index $l^{*} \propto W_0^{*,l^{*}}$.
 
 Finally, calculate $\hat{p}_{\mathcal{M}^{*}}(G, \Gamma)$ via \eqref{eq:normalizingconstantest}.}

 \item{ Step 2: With acceptance probability
 \begin{align*}
 1\wedge \frac{\hat{p}_{\mathcal{M}^{*}}(G, \Gamma)q(\mathcal{M}|\mathcal{M}^{*})}
 {\hat{p}_{\mathcal{M}^{(r-1)}}(G, \Gamma)q(\mathcal{M}^*|\mathcal{M})},
 \end{align*}
 set $\Big( l^{(r)}, \,\mathcal{M}^{(r)},\, \phi_{0:n-1}^{(r),1:N}, \,a_{1:n-1}^{(r),1:N} \Bigr) =\Big( l^*,\, \mathcal{M}^*,\, \phi_{0:n-1}^{*,1:N},\, a_{1:n-1}^{*,1:N}\Big)$. Otherwise, set 
$\Big( l^{(r)},\,\mathcal{M}^{(r)},\,\phi_{0:n-1}^{(r),1:N},\, a_{1:n-1}^{(r),1:N}\Big) = 
\Big(l^{(r-1)}, \,\mathcal{M}^{(r-1)},\,\phi_{0:n-1}^{{(r-1)},1:N},\,a_{1:n-1}^{{(r-1)},1:N}\Big)$.
 
 Return to the beginning of Step 1.}
\end{itemize}
\caption{\label{alg:PMMH}Particle Marginal Metropolis-Hastings (PMMH)}
\end{algorithm}

\end{document}